\documentstyle[12pt]{article}
\begin{document}
\title{    RIGHT   HANDED    WEAK    CURRENTS
IN SUM RULES FOR AXIALVECTOR CONSTANT RENORMALIZATION}
\author{N.B.Shul'gina\\
The Niels Bohr Institute, \\
Blegdamsvej 17, DK-2100 Copenhagen \O,Denmark\thanks{Permanent
address: The Kurchatov Institute,
123182 Moscow, Russia}}
\date{   }
\maketitle
\begin{abstract}
The recent experimental results  on  deep  inelastic  polarized
lepton scattering off proton, deuteron and $^{3}$He together with polari%
zed neutron $\beta $-decay data are analyzed. It is shown that the  problem
of Ellis-Jaffe and Bjorken sum rules deficiency and the neutron  paradox
could be solved simultaneously by assuming the  small  right  handed
current (RHC) admixture in the weak interaction Lagrangian. The possible
RHC impact on
 pion-nucleon $\sigma $-term and Gamow-Teller sum rule for
$(p,n)$ nuclear reactions is pointed out.
\end{abstract}
\medskip
\medskip
\par
In comparing sum rules for axialvector constant renormalization
 with the experimental data one
usually assumes that
the axialvector constant renormalization $\lambda =g_{A}/g_{V}$  is a well
known
value, measured with high accuracy in neutron beta decay. Is it really the
case?
Let us consider the modern experimental status of neutron beta decay in more
details. The axialvector constant renormalization can be extracted either
from the neutron
life time ($ t_{n}$) or from the electron asymmetry ($A$)
according to the well known formulae:
\begin{equation}
f_{n}t_{n}= 2 (ft)_{0-0} /(1+3\lambda ^{2}_{\tau}),
\label{1}
\end{equation}
\begin{equation}
A=-2 \lambda _{c}(\lambda _{c}+1)/(1+3\lambda ^{2}_{c})
\label{2}
\end{equation}
In the Standard Model of the electroweak interaction $\lambda _{\tau}$ should
be equal to $\lambda _{c}$.
 According to the recent  experimen%
tal data on the polarized neutron beta  decay,  which  was  measured
with an accuracy about $10^{-3}$, and the experimental  data  on $0^{+}- 0^{+}$
beta transitions $[1,2],\ \lambda _{\tau}$ and $\lambda _{c}$ differ from each
other at $2.6\sigma $ level.
Indeed, for the electron asymmetry
$A=-0.1126 \pm  0.0011 [3],\ \tau_{n} = 886.7 \pm
1.5 s$ (mean-weighted value of Particle Data Group data [3] and ${\rm re}$%
cently appeared value of ref. [4]) and $ft_{0-0}= 3074.0 \pm  3.5 s\ [1]$:
\par
\begin{equation}
\lambda _{\tau}=- 1.270 \pm  0.002
\label{3}
\end{equation}
\begin{equation}
\lambda _{c}= -1.257 \pm  0.003 \label {4}
\end{equation}
\par
 In the papers [5,6] the neutron paradox was
explained in the framework of the  left-right  symmetric
model $SU(2)_{L}\times SU(2)_{R}\times U(1)\  [7]$ and the possibility of the
 RHC admixture in the neutron beta decay was pointed out. It should be
 stressed that the vector  and the  axial-%
vector constants are "renormalized" by RHC in a different way, so
that the "nucleonic" or "bare"  axialvector constant renormalization
 $\lambda _{N}$
may differ significantly from well known value $\lambda =-1.26\  [5,6]$.
\par
In the the simplest manifestly left-right  symmetric  model  of
$SU(2)_{L} \times SU(2)_{R}\times U(1) $  the expressions for $\lambda _{\tau}$
  and  beta  decay
asymmetry $ A$  have the following form [5]:
\begin{eqnarray}
\lambda _{\tau}= \lambda _{N}[(Z+X)/(Z-X)]^{1/2},\label{5}\\
A = -2\lambda _{N}(\lambda _{N}+1+Y\lambda _{N})
( 1-X^{2})^{1/2}/[(Z-X)(1+3\lambda ^{2}_{\tau})]
\label{6}
\end{eqnarray}
\noindent where
\par
$$
X = \sin  2\zeta  , Y= (1-\eta ) \sin  2\zeta /(1+\eta ) ,
$$
\begin{equation}
Z=(1+\eta ^{2})/(1-\eta ^{2})
\label{7}
\end{equation}
Both  experimental values are functions of  the  model  parameters,
 $\eta ,\zeta ,\lambda _{N}$,  which  have  the  following  physical  meaning:
$\eta  =
(M_{1}/M_{2})^{2}$ denotes the squared mass ratio of the $W_{1}$and $W_{2}$
 bosons; $\zeta $ - the
mixing angle of these bosons $(W_{L}= W_{1}\cos \zeta  + W_{2}\sin \zeta ,
 W_{R}= - W_{1}\sin \zeta  +
W_{2}\cos \zeta ); \lambda _{N}$ - the bare relative renormalization of the
 axialvector nucleon current. It is important to stress here that
 the $\lambda_{N}$ in Eq.(5-6)
is the value, which should be determined in a way independent of the nature of
weak interaction e.g.,from Adler-Weisberger sum rule.
\par
In the $SU(2)_{L}\times U(1)$ limit the parameters tend to :
\begin{equation}
\eta  = 0 , \zeta  = 0 , \lambda _{N} = \lambda _{\tau} = \lambda _{c}
\label{8}
\end{equation}
\par
The set of the experimental data on the $\lambda _{\tau}, A, \mu $-decay [8]
enables
to restrict the range of permissible values of  the
$\lambda _{N},\zeta ,\eta $  parameters.
\par
$$
0.003 \le  \zeta  \le  0.054 , \eta  \le  0.036 , -1.265\le  \lambda _{N}
\le -1.131
$$
$$
-0.054 \le  \zeta  \le  -0.020 ,\eta  \le  0.024 , -1.415 \le
\lambda _{N}\le  -1.20
$$
\begin{equation}
 M_{WR} \ge  427 GeV ( 98\% {c.l.}) \label{9}
\end{equation}
\par
\noindent As is seen from Eq.(9), in each of two regions, $\lambda _{N}$
 varies  within
$10\%$ interval. The muon data [8] does not
indicate the RHC and is used here in order to restrict upper limits of
 the $\zeta$ and $\eta$
parameters, while the neutron data [3,4] provides the lower limits.
In principle, the muon data can be used only if the muon-electron
universality is assumed.
\par
The nonzero contribution of RHC in nuclear beta decay was  con%
firmed also by the joint analysis of neutron  and $^{19}Ne$  beta  decay
[9]. It is necessary to notice that in the simplest left-
 right symmetric model the large mixing angles ($\zeta\ge 0.006$) disagree
with the unitarity of the Cabbibo-Kobayashi-Maskawa matrix for three quark
generations. However this can be avoided in an extended version of the model.
Anyway, the unitarity problem is a theoretical one, while in this
 paper only experimental data will be analyzed.
\par
{\bf Ellis-Jaffe sum rule}
\par
An interesting aspect of the RHC arises when considering  the  EMC
experimental data on deep inelastic polarized lepton- proton scatte%
ring. Let us recall the essence of the problem. EMC measurements  of the
spin dependent proton structure function [10]:
\begin{equation}
 {\Gamma ^{p}} {=}\int_{0}^{1} g^{p}(x)dx {=} {0.126} {\pm } {0.010} {\pm }
  {0.015}\label{10}
\end{equation}
\noindent indicated a significant deviation from Ellis-Jaffe sum rule  [11],
which was derived using $SU(3)$ current algebra with the assumption of
unpolarized strange quark sea:
\begin{equation}
 {\Gamma ^{p}} {=}\int_{0}^{1} g^{p} (x)dx {=} {{1}\over{12}}\Biggl
 | {g_{A}\over g
_{V}
}
\Biggl | {\Biggl ( 1} {+ {5\over 3}} {{3F-D\over F+D}\Biggl )} \label{11}
\end{equation}
\noindent where $F$ and $D$ are the $SU(3)$ invariant matrix elements of
 the  axial
current, which are usually deduced as a weighted mean values of  all
types of hyperon semileptonic beta decay data.
\par
After correcting for the QCD radiative effects  [12]  the  integral
(11) becomes [10]:
\begin{equation}
\Gamma ^{p}= 0.189\pm  0.005,
\label{12}
\end{equation}
where the old value $g_{A}/g_{V} = 1.254\pm  0.006$ from  neutron  beta  decay
and $F/D = 0.632\pm 0.024$ from overall hyperon beta decay fit were used.
The discrepancy (more than two standard deviations) between predic%
ted (12) and experimental (10) values,  known as the  "proton  spin
crisis", created a lot of theoretical  explanations  (  see,  for
example, the review [13] and references therein).  In  the  most  of
them the proton spin is supposed to be carried by gluons or  orbital
angular motion. However these models can't, probably, explain polarized pro%
ton - nucleon data at high energies (for details see ref.[14]).
\par
The RHC impact was not considered as  a  possible  explanation
although this seems to be simplest. Indeed, $g_{A}/g_{V}$  in the Ellis-  Jaffe
sum rule is just the same $\lambda _{N}$ value considered above. Measured
 values, say $\lambda _{\tau}$, can be affected by RHC as  seen from  Eq.(5).
 As  is
mentioned above, $\lambda _{N}$, in principle, should not be equal
 to $\lambda _{\tau}$ or $\lambda _{c}$,
extracted from the neutron life time or the electron  asymmetry,  especially
if one takes into account the $2.6\sigma $ discrepancy between
 $\lambda _{\tau}$ and $\lambda _{c}$. If one supposes that the RHC do
exist in  the neutron
beta decay,  the sum rules  in  deep  inelastic  lepton-
nucleon scattering should be tested in another way.  First  of  all,
one should realize that the $g_{A}/g_{V}$ value in Eq. (11) is neither
 $\lambda _{\tau}$ nor $\lambda _{c}$
nor their mean value, but should be taken from Eq. (5). Secondly, the
$F/D$ value can differ from the least square fit to all hyperon decay data,
which can be affected by RHC in  different ways.  Strictly  speaking,  for a
precise estimation of the $F/D$ value all hyperon decay data should be
revised in the framework of $SU(2)_{L}\times SU(2)_{R}\times U(1)$  model.
 A thorough
analysis of hyperon $\beta $ decays for the case of RHC
 is tedious and will be  done  elsewhere.  Perhaps  the
shortest and rather accurate approach to the problem is  to  rewrite
Eq.(11) in the following form:
\par
\begin{equation}
 {\Gamma ^{p}}{=}\Biggl \{
{2\over 9}\Biggl | {g_{A}\over g_{V}}\Biggl | ^{np}
{-} {{5\over 18}\Biggl | {g_{A}\over g_{V}}\Biggl |^{\Sigma n}
\Biggl \}\Biggl [
(Z-X)/(Z+X)\Biggl ]^{1/2}} \label{13}
\end{equation}
\noindent where indices $np$ and $\Sigma n$ mean that the ratio relates to
the neutron $\beta $- decay and
to $\Sigma ^{-}\rightarrow  n + e^{-} + \bar{\nu }$ decay,
 respectively. Eq.(13) can be easily derived using  the SU(3) matrix elements
for neutron and $\Sigma ^{-} $- hyperon $\beta$- decay:
$\left({g_{A}}/{g_{V}} \right)^{np} = F + D$;
$\left({g_{A}}/{g_{V}}\right)^{\Sigma n} = D - F$.
\par Two high statistic experiments [15,16] with unpolarized  beams  give
 $\mid {g_{A}/ g_{V}}\mid ^{\Sigma n}$ $= 0.36 \pm  0.04.$ Using this  value,
  $\mid {g_{
A}/ g_{V}}\mid ^{np} = \mid \lambda _{\tau}\mid  = 1.270
\pm $ 0.002 and correcting for QCD radiative effects one can obtain:
\par
$$
{\Gamma ^{p}=(0.170\pm 0.010)} {[(Z-X)/(Z+X)]}^ {1/2}
$$
When the RHC parameter $\zeta $ varies within the interval
 $0.022\le  \zeta \le  0.054,$
which is permitted by neutron and muon beta decay data, the new  version
of Ellis- Jaffe sum rule (13) agrees with the  experimental
data.
\par
{\bf Bjorken sum rule}
\par
A more fundamental sum rule for deep inelastic polarized  lepton-
nucleon scattering is the Bjorken sum rule, since $SU(3)$ symmetry is  not
assumed. This sum rule relates the integral over $x$ of the difference
of neutron and proton structure functions to  the  bare  axialvector
constant renormalization in the following way:
\begin{equation}
{\int^{1}_{0}} {(g^{p}(x)-g^{n}(x))} {dx} {=} {1\over 6}\Biggl |
 {g_{A}\over
g_{V}}\Biggl | \label{14}
\end{equation}
\noindent So an observation of the spin dependent structure function
  of a  neutron
could be the less ambiguous way to estimate the bare  axialvector  cons%
tant renormalization.
The neutron spin structure function was determined very recently
by measuring the asymmetry in deep inelastic scattering of polarized
electrons from a polarized $^{3}$He target [17]. Experimental $\Gamma ^{n}$
 value
is found to be $- 0.22\pm  0.11.$ Together with proton structure function
(11), corrected for relevant $Q^{2}$, and Bjorken  sum  rule  (14)  it
gives :
\par
$$
0.833 \le  \mid \lambda _{N}\mid  \le  1.187 (68\%{ c.l.})
$$
\noindent (with the mean value being equal to unity!).
In terms of RHC parameters it means:
\par
$$
0.03 \le  \zeta  \le  0.17
$$
\noindent which does not contradict to neutron and muon data.
\par
Measurements, carried out with an  accuracy  better  than $10\%$,
could throw a light on the nature of Ellis- Jaffe  and  Bjorken  sum
rule violation. In any case, even for more  precise  future  experi%
ments a reasonable difference ( $\approx 10\%)$  between  axialvector
constant renormalization, extracted from the Bjorken sum rule, and  that
from the neutron lifetime can take place and can be explained  in  terms
of RHC.
\par
{\bf  Pion-nucleon $\sigma$- term}
\par
  The experimental
value of the pion decay constant  $F_{\pi }$, deduced from the weak
 pion decay:
$\pi  \rightarrow  \mu + \nu _{\mu }$ could be also renormalized by RHC. Indeed
 in the Standard model the pion decay is helicity suppressed and takes place
  only  due  to
the muon mass. If the admixture of RHC is allowed, the decay probability
increases. This means that the experimental value  $F_{\pi \exp }$ is greater
than bare one, $F_{\pi B}$.
\par
After this remark, let us reanalyze the pion-nucleon $\sigma $  term.
Along with EMC experimental data on deep inelastic  polarized  muon-
proton scattering, a large experimental value of the pion- nucleon
 $\sigma $-term
is considered as an evidence for large strange quark content of pro%
ton [18]. In the light of $F_{\pi }$ renormalization by  RHC  the  surprisingly
large experimental value of the pion-nucleon $\sigma $ term,  as  compared
with
the theoretical one under assumption of zero strange quark sea component
of proton, can be explained.  Keeping  aside  the
details of calculations, as well an experimental uncertainties, ana%
lyzed thoroughly in Ref.[19], let us consider  the  treat%
ment of experimental data. Actually  ( for details see the review of
E.Reya [20]) one can extract from experimental data only the  ratio:
$\sigma _{\pi N}/ F^{2}_{\pi B}$. If one assumes that $F_{\pi }$ is
renormalized by RHC in the  same
way  as $\lambda_{N}$ (that is true for the simplest version of the
 left-right  sym%
metric model), then
\par
$$
F_{\pi B}\approx  F_{\pi \exp } (1-2\zeta )
$$
\noindent and the experimental values $\sigma _{\pi N} = 45\pm 10$ MeV [21]
 can be
reconciled  with
theoretical values $\sigma _{\pi N}= 23 \pm  5 [19]$ within $1\sigma $ interval
 for $\zeta  \approx  0.05.$,
  a  value  which
explains also the Ellis-Jaffe sum rule deficiency, as well as the neutron
 paradox.
\vspace{16mm}
\par
{\bf Gamow-Teller sum rule for nuclear reactions}
\par
There is a well known experimental fact [ 22 ], that in $(p,n)$
 reactions
on nuclei
the Gamow-Teller sum rule [23]
\begin{equation}
  S^{+}(GT)-S^{-}(GT) = 3\lambda^{2}{ (N-Z)}\label{15}
\end{equation}
is not exhausted at low  excitation energies including the Gamow-Teller
giant resonance  when one uses ${\lambda = \lambda_{\tau}}$. The deficiency
 or, so called "quenching", is about $40\%$. The sum rule (15) is model
 independent if non nucleonic degrees of freedom are not introduced and
 isospin is
a perfect symmetry.
Different aspects of the nuclear structure as well as delta-isobar excitations
were intensively discussed as a possible explanation of the quenching effect
[24].
But if one takes into account that in the strong  processes
$\lambda $ in Eq.(15) is the bare value and
should not be taken from neutron life time one half of the  missing strength
can be explained by the $\zeta\approx 0.05$, a value
which explains the Ellis Jaffe  and Bjorken sum rules deficiency as well as
the  pion-nucleon $\sigma$ - term and the neutron paradox. If RHC do exist
one can obtain the relation:
\begin{equation}
 [S^{+}(GT)-S^{-}(GT)]_{strong}/[ S^{+}(GT)-S^{-}(GT)]_{weak}\approx (1-4\zeta)
 \label{16}
\end{equation}
which is independent of nuclear structure models and enables, in principle,
to deduce the RHC
parameter $\zeta$ by comparing the "strong" and "weak" experimental
GT strengths in the same region of excitation energies. Of course,
 the experimental uncertainties have to be
at least less than $20\%$. It should be noticed here, that in contrast
to Eq.(16),
the procedure of the GT strength extraction from the (p,n) reaction
 cross sections
involves nuclear structure parameters. Therefore it would be worthy
to test the prediction (16) in
lightest nuclei, where the nuclear structure is calculated more reliably.
\par
The pure $\lambda _{N}$ can be deduced from the experimental data, if one
  uses  the
Adler- Weisberger sum rule [25,26]:
\par
$$
{1} {-} {1\over\lambda_{N}^{2} } {=} {4 M\over g_{\pi nn}^{2}} {1\over \pi }
{\int_{
M_{N}+m_{\pi}}^{\infty }{W dW\over W^{2}- M_{N}^{2}}{[}{\sigma ^{+}_{0}(W)}
{-} {\sigma ^{-}_{0}(W)} {]}}
$$
\noindent where $\sigma ^{\pm }_{0}$ is the total cross section for
 scattering of a zero mass $\pi ^{\pm }$  on
a proton at the center of mass energy W.
\par
 Up to now, only two estimations of the $\lambda_{N}$ value
 from this relation exist: one, given by  Adler
$[25] -\lambda _{N}= -1.24$ and another, given by Weisberger $[26] -
\lambda_{N}= -1.15.$
Both estimations were done in 1965. Since that time new experimental
data on pion -proton scattering has appeared and strong  interaction
constant $g_{\pi nn} $, has been revised. Therefore it would be  worthwhile to
reanalyze Adler- Weisberger sum rule in order to extract bare axial%
vector constant renormalization more accurately.
\par
The problem  of
 the bare value of the axialvector constant renormalization is important
also for calculations of the counting rates in solar neutrino detectors,
 which employ
(p,n) experimental data for weak process calculations (see, e.g. [27]).
\par
In conclusion, it should be emphasized, that intimate connecti%
on between low energy weak processes and high energy scattering pro%
cesses, based on current algebra, provides sensitive tests for the Stan%
dard model of strong and electroweak interactions.
\par
\medskip
I am grateful to Yu.Gaponov, P.Herczeg, W.Weise, C.Gaarde, I.Towner and J.Bang
 for  useful
discussions. I am also thankful for the hospitality of The Niels Bohr
Institute and NORDITA where part of this work has been done.

\end{document}